# Energy Experience Design

Brian Sutherland
University of Toronto
Toronto, Ontario Canada
b.sutherland@utoronto.ca

## ABSTRACT

The material footprint of information and communications technology (ICT) systems is both significant and growing, inspiring a variety of conversations around sustainability and climate justice. In part this effort has been catalysed by past scholarship and analysis from the LIMITS community. This paper examines energy storage systems for computing, particularly batteries -- which are discarded at the rate of 15 billion a year worldwide.  The International Energy Agency (IEA) is now referring to the energy transition toward low carbon systems as a critical mineral problem, and countries are speaking openly of 'mineral security' in policy documents. In this paper I 1) present a definition for energy experience and what this means for the design and making of devices, interactions and experiences. I also 2) explore a series of electronics device prototypes converted to run from batteryless sustainable energy that are extremely long lasting, and make limited use of critical minerals. As transitional energy experience device-design experiments, what do prototypes like these suggest for more mainstream, mass-manufactured systems?

## .1   Absences in Material Computing

Past LIMITS papers have discussed the requirements of computing and networks, limits awareness [1], natural resource dependency [2] and material consumption with respect to values [3]. A variety of response 'transitional' systems involving LIMITS concepts have been presented in previous years: such as solar-powered workstations [4], servers [5] and server-networks [6].

As I discussed in a previous paper about degrowth, [4] ICT devices take significant energy and materials to manufacture, much more than in their use -- so, their full material costs over their lifecycle need to be valued appropriately in market representations. This appropriateness does not mean pricing devices so they are only available to the wealthy; rather it means obtaining greater collective value out of their social investment in making -- by extending their use before they are discarded, increasing their durability, improving their repair, trade, and upcycling potential, and using them more intensively and in shared ways when they are new. Progress has been made in this direction with right to repair legislation, such as Britain's 2021 law to require manufacturers to carry 10 year inventories of spare parts [7]. Less progress has been made in my province of Ontario with the successive defeats of Right to Repair legislation Bills 72 and 187. Design as well as legislation has a role to play in these regulatory conversations in that products that are available in markets have specific features and use-values that advocate for approaches and policies toward everyday living.  Consider for example myriad AC adapters which translate mains electricity into discrete direct current for individual devices, compared with the more modern universal serial bus (USB) format that can power any portable product. Just as a conversation needs to occur around repair, a similar shift needs to occur in media around standards of manufacture which support interoperability and simple repair and maintenance.

A related concept to the search for edges, boundaries, or limits is to cast around for absences. Feminist scholar Sara Ahmed suggests: "we need to understand how it is that the world takes shape by restricting the forms in which we gather" [8]. While Ahmed was speaking about diversity and representation in citational structures of knowledge in academic disciplines, this argument easily extends to gatherings of objects with cultural significance -- where excluding them, regulating their supply or distributing and pricing them in certain ways within markets enacts scripts that shape certain futures and exclude others. Aside from repair and interoperability, designs inculcate specific experiences involving duties of care and energy, where cultural knowledge of their operation surrounds or circumscribes the range of options we might consider in a particular future.

One particular future being contemplated in the energy transition involves the need for 'critical minerals', which the IEA defines as rare materials required to address energy storage and intermittency in a sustainable low-carbon (electric) culture. Critical minerals are particularly needed for batteries: "lithium, nickel, cobalt, manganese and graphite are crucial to battery performance, longevity and energy density."[9] A significant matter of concern is how to manufacture enough rechargeable batteries in a low carbon way to reduce climate change without creating more of it, while ensuring that communities whose areas are disrupted by extraction of minerals benefit fairly rather than being exploited, the concept of "reciprocity". But in the words of Ezio Manzini, the academic who wrote about design and environmentalism in the 1990s, sometimes "the problem is not so much one of evoking an ethical principle based on an environmental necessity, as of proposing solutions that appear to be better than those presently in use." [10].  This article looks at certain absences in energy experience design and computing.

## 2.   Energy Experience Design

I first heard Tega Brain, Benadetta Piantella and Alex Nathanson of New York University use the expression "energy experience design" in conversation, to describe one of their technology projects: Solar Protocol. Interestingly, it's also a job title at the electric vehicle maker Tesla Motors [11].  At the time of writing, driving from Toronto to Montreal in an electric vehicle requires both rapid recharging and an available charge-unit on the route, since the distance is beyond the range of most vehicles. Just as user experience designers improve the experiences people go through when interacting with companies, services and products, energy experience design focuses on social experiences involving energy design, with some methods in common with anthropology.

The concept of energy experience design is perhaps to acknowledge the role of the user in the configuration of sustaining devices, but shift the field of view slightly away from what users want, toward the socially mediated, material, provisioning, earthbound problems of the journey.  To consider, for example, the influence of objects like sweaters in the energy relation and behaviours involving home heating, but in the realm of information and communications technology. Like in actor-network theory, where objects "operate to stiffen, to reorganize, or to dissolve what we normally think of as social relations" [12] energy provision and use systems have a framing effect both to how we act with our devices in the moment, and our cultural behaviours around device-use over a longer period.  Cornucopia networks which support manufactured computing devices and electrical storage can make information systems easy to take for granted, simple to throw out and replace, creating unsustainable energy experience behaviours in aggregate.

Consider for example, batteries, which are integral to the experience of portable computing. Batteries have had standard sizes for interoperability between companies and their products since 1898, when D cells were introduced. We might reflect on common energy experiences of using batteries: that they last a limited amount of time and need to be replaced, they often leak chemicals which damage the electronics products in which they are situated, and they are repetitively consumed on a massive scale: 15 billion units a year worldwide. A germane energy experience question is: could energy storage work better? It is clear from the history of batteries in consumer products, that some varieties we commonly consider to be single-use now are actually rechargeable. The manual of a Hoffman 709 solar radio from the early 1960s, for example, discusses a circuit to recharge single use batteries from the solar panel. A magazine article in July 1967 Popular Electronics similarly reviews the recharging of dry cell batteries [13]. Single use alkaline batteries used to have a recharger you could buy, and a recent project called 'The Regen Box' that "regenerates your single use batteries" has resurrected this capability from the expired patent.  They claim 25 recharges are typically possible [14]. Aside from single use battery chemistries being actually rechargeable, batteries specifically designed for recharging have been around for decades, and we might ask (at least in North American stores) why aren't they easier to find? It costs as much to manufacture a single use battery as a rechargeable battery [15], yet the social and end-user cost is significantly more from single-use batteries, in the full life cycle analysis. In a more modern example we might consider electronics devices that have overwritten the experience of universal interchangeable batteries -- their energy storage comes only from the manufacturer and is often parts-paired with the product-year combination for which it was designed.  Compare this energy storage logic with an early Hoffman Solaradio from the 1950s which came with a battery adapter, supporting not just one but TWO universal sizes:  C or AA batteries!  Are OEM product-paired batteries a reaction to poor batteries in the market, an attempt to improve size, weight and performance for products? Or are they an obsolescence strategy, in as much as ICT products attract consumer dissatisfaction when they cease to store electricity and constantly need recharging. Product-examples existed in the past which provided for more sustainable energy experiences.  The question, then, is how future products might be designed to reduce repetitive consumption around energy storage.

## 3.   Transitional Devices

A few modern electronics devices reflect a more low carbon and sustainable frame involving energy storage.  One is Lukas Hartmann's MNT Reform Laptop, where the batteries are lithium iron phosphate (don't use cobalt, a critical mineral) [16]. These batteries are also user exchangeable, easily accessible within the case and use a standard rechargeable size known as 18650 - 18 mm diameter, 65 meter height and 0, cylindrical shape.  18650 batteries are the building blocks of many rechargeable products, and a further benefit of lithium iron phosphate battery chemistry is that not only do they last longer (recharge more times), they are safer -- no thermal runaway is possible as with lithium ion batteries. The battery energy storage strategy of the MNT Reform laptop, therefore, reflects a commitment to reform around open content: it is user-repairable from standards based universal interchange units.

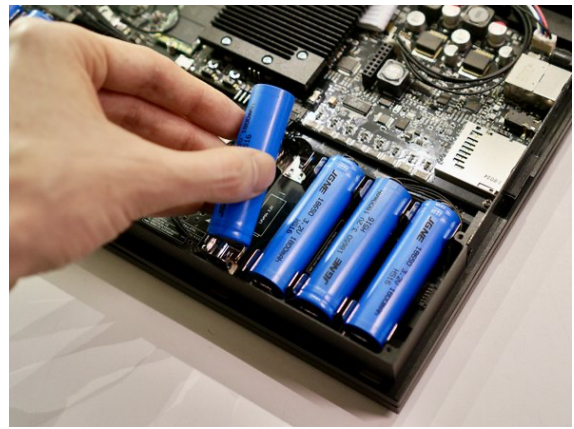

**Figure 1: MNT Reform Laptop, LiFePO4 18650 replaceable batteries (CC-BY-SA 2.0)**

Cell phone markets also have a number of 'reform' devices: sustainable entries by way of energy storage. At the time of writing, Samsung has two Android phones in the Galaxy XCover series which have user-replaceable batteries, likewise Nokia the C21 and C100 and there's also the Fairphone . Fairphone is somewhat a reaction to market absence, like the MNT Reform laptop -- it's a phone that comes with a screwdriver: easy to open, easy to repair, and easy to maintain, where parts and upgrades are guaranteed for five years. The battery clips out relatively easily, unlike other smartphones where a battery would need to be removed by taking most of the phone apart with a licensed repairer.  The Fairphone company website features detailed video instructions. Aside from its repair and upgrade strategy, Fairphone recycles a phone for every one sold [17]. So in these few products there are some new market framing relations, innovations with respect to making batteries work better and in a more sustainable way. But is that enough? While Lucy Suchman speaks of design as "responsible practice ... characterized by humility rather than hubris, aspiring not to massive change or discontinuous innovation but to modest interventions" [18], I wonder whether more deliberate breaks with past technologies aren't occasionally warranted. In 1900 38% of automobiles in the US were powered by electricity, only 22% by gasoline. Flooding markets with massive quantities of low cost manufactured vehicles of the gasoline type, the Model T Ford, redefined the network of relations around transportation energy experience, infrastructure, and global warming.

## 4.  Alternative Storage

While there are degrees of sustainability within battery designs, we might consider early alternative experiences with electricity storage, i.e. non-batteries.  One was the Leyden jar, an electricity storage device with layers of metal foil separated by glass, the forerunner of the device we now call a condenser or capacitor. Capacitors are relatively simple devices: they store electricity as physical static charge on alternating layers of conductive and nonconductive material, rather than as chemical changes which may or may not be reversible.   Capacitors have many uses in circuits. An early Bell Laboratories book on relays for example, discusses why telegraph relays could actuate a million times in a day without wearing out compared to telephone relays [19]: capacitors were used as 'spark killers' for these relays, reducing the material impact of switching large currents on and off.  Just like we can choose to make relays last longer by designing and making them differently, so too is it possible to make electricity storage last longer and less the subject of repetitive consumption. Before discussing types, I'll observe that 'capacitance' or ability to store charge is measured in Farads, after physicist Michael Faraday. Typical capacitance is in milli, micro or nano Farads and not much electricity is stored in most capacitors, usually, compared with batteries.  Experiences with these types of capacitor are perhaps the origin of a lot of wrong ideas about storage capacitors, such as: they don't store charge in useful quantities or for a useful period of time, like batteries do.

In 1954, which was coincidentally the year photovoltaic or solar cells were invited, Howard Becker at General Electric invented the 'Low Voltage Electrolytic Capacitor' which could support capacitances between 0.8 Farads up to 6 Farads, as described in US Patent 2,800,616 [20]. One conclusion he mentions in the patent is that the capacitor can be "left on short circuit for an indefinite period of time without significant damage or breakdown of its qualities". Practically speaking that translates to the ability to store electricity and discharge and recharge a million times, compared with a few thousand times for the best rechargeable batteries. Capacitors get *soldered on* to electronics boards because they last decades, like the other 'permanent' components.

Since 1954 and the creation of Becker's low voltage 'supercapacitors', we've seen the development of 'ultracapacitors', capacitors in the thousands of Farads range, about the size of a juice can.  At one point it was thought that ultracapacitors might power electric vehicles or EVs because of their ability to store, charge and discharge electricity effectively, discharge and recharge rapidly (much faster than batteries) and of course their durability with respect to cyclical charging.  Ultracapacitors are used for accelerating and braking in high performance race cars and for passenger buses. A Scientific American article in 2007 discussed a few of these applications, calling ultracapacitors 'The Dark Horse in the Race to Power Hybrid Cars' [21]. More importantly to the application of everyday consumer electronics though, a variety of mass-produced products in the 1980s and 1990s used supercapacitors instead of batteries, specifically the solar powered Citizen Eco-Drive and Swatch Solar watches. These timepiece designs took advantage of the Panasonic Gold Cap, a small supercapacitor alternative electricity store designed in the aspect of a watch battery [22].  Ironically, one of the peculiar problems of decades-old Citizen Eco-Drive watches is that owners believe they need repair when all that is required is to put them in the sun to recharge the capacitor to continue working. Alternative storage products which use energy harvesting in combination with capacitors, instead of batteries are so durable in fact, they defy the energy experiences of people who expect to have to replace a battery.

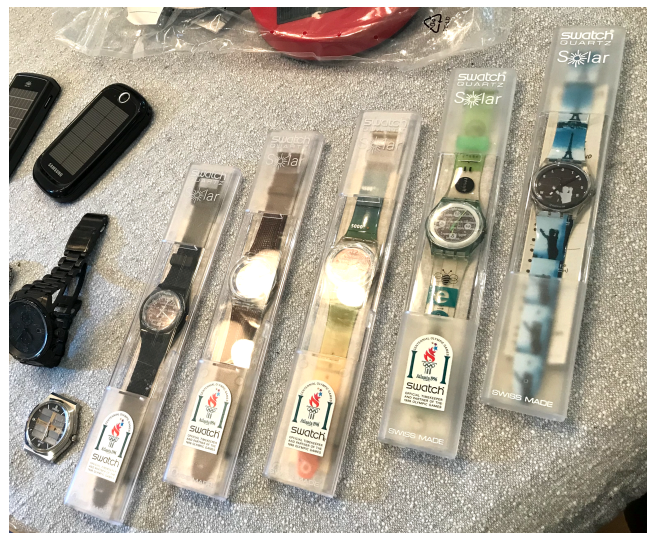

**Figure 2: Swatch Solar watches produced during the 1990s** -

In previous years I described a transitional energy harvesting computing system: a prototype Raspberry Pi 4 desktop computer powered by a solar panel and buffered by 24 x Maxwell D-Type 350F supercapacitors, each rated for 2.7V, balanced with a protection board in four groups of six. The monitor was a 11W 27" HP flat panel. This resulted in a solar powered, batteryless workstation.  In 2020 a new kind of capacitor came out, called the 'hybrid supercapacitor' initially from Eaton Electronics in 220F size, charging to 3.8V and discharging to 2.2V a voltage more amenable to digital logic. It is similar in function to ultracapacitors except the energy density is eight times higher, reducing the juice can to the size of a D cell for a similar amount of charge. This in effect means comparable performance to batteries with respect to the amount of volume required per unit of electricity, the energy density, but with significant advantages: fewer critical minerals and 500 000 x recharges [23]! I used these to create a transitional device, adapting an open source battery FM Radio in a mason jar to give it 12 of play from four 220F Eaton hybrid supercapacitors wired in parallel, recharging from a small solar panel.

In 2022 cda of Taiwan came out with 1100F 4V hybrid supercapacitors, five times more capacity than Eaton's 220F product, for approximately the same price. In Figure 3 I show the modified FM radio, which now gives 75 hours of play and recharges daily on the windowsill. Since building this radio, I have tested the alternative storage on other household devices. For example, in Figure 4 I show a test circuit where the purple cda 1100F hybrid supercapacitor discharges through a high wattage LED light that shines brightly for about two hours. While the system for this bike headlight is not substantially different from a USB rechargeable lithium ion battery flashlight, materially and in durability it is advantageous to the user and the environment. Furthermore, with only minor modification, a twenty-cent TP4056 linear li-ion battery charging circuit can be used to charge the hybrid supercapacitor from a standard USB port, stopping when it's "full".

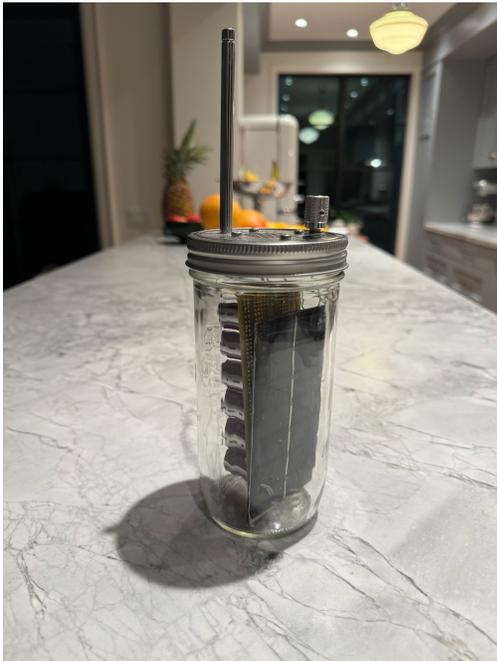

**Figure 3: Open electronics FM radio in a mason jar radio with 6600F of hybrid supercapacitors in parallel - 75h of play**

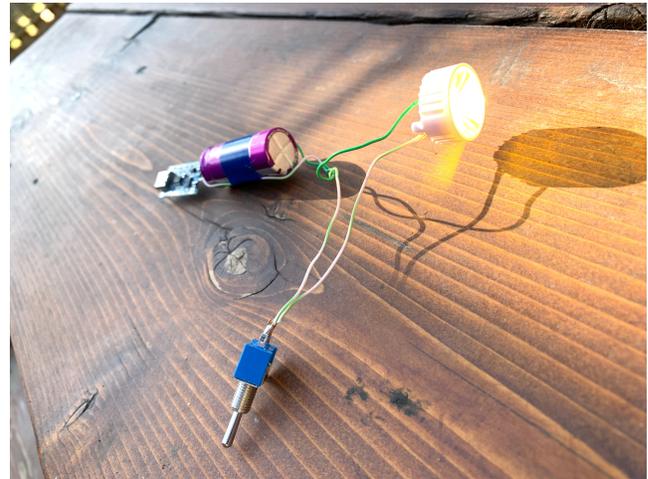

**Figure 4: 1W LED bike headlight, powered by a cda 1100F hybrid supercapacitor**

Based on these two device-tests, it is clear that large hybrid supercapacitors have enough electricity storage to buffer portable digital consumer energy systems effectively and for significant periods as a sustainable alternative to batteries, while offering a similar energy experience.  This approach has some potential to address the problem of the high consumption of batteries and the need for critical minerals in a positive way.

In the case of the radio, leaving the mason jar in the window keeps the capacitors charged up, and most days it's only12 hours until the sun appears.  The portable bike headlight similarly recharges outside an office building on campus where my bike is locked for the workday.   But can these alternative electricity storage power strategies support more serious computing device use, while providing reasonable energy experiences?

## 5. Transitional Computing

For a first test platform I selected a discarded Acer One netbook from eWaste, removing the power adapter so that the wiring was exposed. This computer operates ordinarily from a 19V 0.8 amp DC supply. I connected this netbook to an array of seven 1100F 4V hybrid supercapacitors in series which could make 28V, mediated by a buck/boost power regulator to reduce the voltage down to a relatively steady 19V. The intention was to use this with a portable folding solar panel as a lightweight batteryless mobile computing system that could operate continuously from a nearby window or outdoors. This did, in terms of experience, however, require that I avoid windowless spaces and adjust my user behaviour to accommodate the power usage requirements and processing capabilities of the netbook.  There was also a thoughtful management of energy storage required to achieve specific goals, such as scheduling webinar-meetings for periods of significant charge accumulation.

While the Acer laptop system did boot and work, the difficulty of maintaining a significant energy flow from low voltage capacitors suggested some further work in the power conditioning area needs to occur before this strategy can be used with higher power draw devices.  I also found the hardware slow to use for tasks I was accustomed to doing.

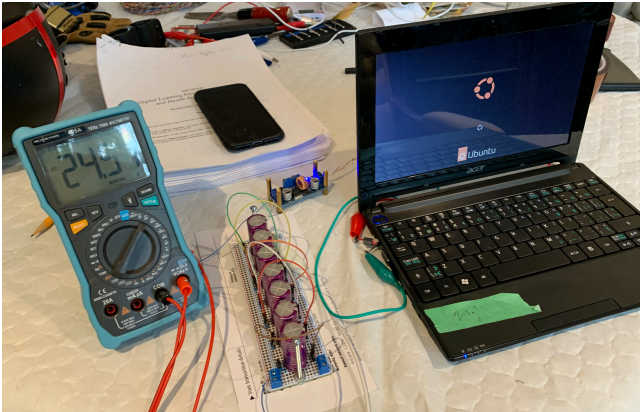

**Figure 5: Acer netbook powered by hybrid supercapacitors**

So for a second test subject I used a laptop that was relatively new, a 16" Macbook Pro from 2021 with an ARM M1 Pro chip marketed under the name "Apple Silicon".  This laptop charges from a USB-C connection, provided there is sufficient current.  To support the 5 volts required by the USB-C connection standard I used a cheap (less than $1) USB converter that I wired to a 12V, 20W panel.   On a bright sunny day, the panel supported active charging of the laptop (Figure 6), slowing if not stopping the battery level from declining over a period of several hours of use.  In partially cloudy conditions, the laptop actively switched between charging and not-charging with each passing cloud, and the battery charge did decline, but slower than without the panel.  The experiment suggested that laptops may be on the threshold of low power and energy conditioning operation that could be more routinely supported by solar panels.   As the test occurred on Apple hardware, Apple might consider reviving the many patents taken out for integrated solar powered computing devices in 2010, beginning to actively produce devices which employ this strategy.

## 6. Discussion

Battery induced obsolescence and the rapid-cycling of mass-manufactured consumer electronics reflect market logics that value repetitive consumption, turning vast quantities of raw materials quickly into new devices and generating atmospheric carbon.  In a future with limits this cycling is unsustainable.  More straightforward repair and upgrades, with a strategy of energy harvesting and alternative electricity storage, as has been used in the Citizen Eco-Drive watches for decades would support devices much more sustainably and for longer periods of time.  This assertion is supported by a number of historical device capabilities in the development of consumer electronics as well as some modern prototypes which I have presented here.  Mass manufactured devices are not inevitable: alternatives exist that could renormalize and reframe the experience of information and communications technology around less consumption.  This depends, however, on a significant committment to alternatives: Wright's experience curve, where manufacturers learn to make sustainable devices more cheaply, needs sufficient time to work. I suggest therefore, that to move the energy transition forward requires networks of open scholarship and open electronics manufacture supported by responsible and collaborative networks of private and public innovation, a sort of community coalition of low-carbon design, in which energy experience figures significantly.

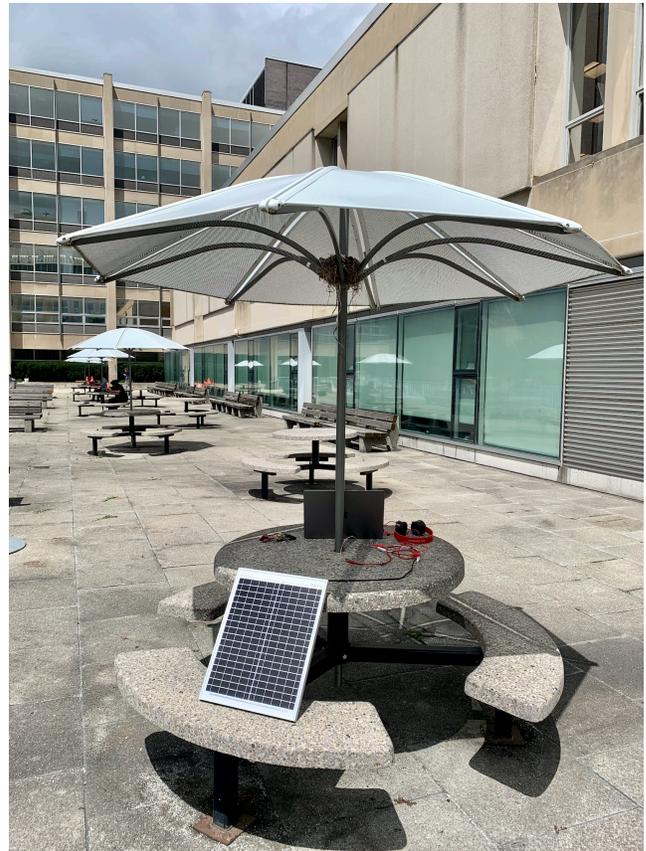

**Figure 6: Arts & Science patio, University of Toronto Macbook Pro 16" M1 Pro, powered by 20W solar panel**

## 7. Conclusions

Energy experience design, particularly with respect to electricity storage is a useful concept to explore in that it refocusses analysis on how we design systems to meet sustainability goals and the signfiicant problems created by user behaviour in repetitively purchasing and discarding consumer electronics.  In the context of ambient available energy as well as the long-term durability of systems, the planetary impacts of manufacture, these proposed energy harvesting batteryless systems are consistent with consuming less and valuing materials and manufacture more substantially, while recognizing many recent and innovative technology advancements in the electricity storage field.  In the market absences of these products exist opportunities to make computing a less consumption oriented enterprise, both in embodied manufacture and during operation, as part of a serious committment to degrowth, recognizing this most useful of tools and its integral relation to human activities.

## ACKNOWLEDGMENTS

Thanks to my spouse Monique for her support, the University of Toronto for access to research resources and computing devices in support of the academic work, and my LIMITS community reviewers for the helpful feedback on improving this article.

## REFERENCES

[1] J. Chen, "A Strategy for Limits-aware Computing," presented at the LIMITS '16, Irvine, CA, USA, Irvine, CA, USA, Jun. 2016. doi: http://dx.doi.org/10.1145/2926676.2926692.

[2] B. Raghavan and S. Hasan, "Macroscopically Sustainable Networking: On Internet Quines," in LIMITS 2016, Irvine, CA, USA, Jun. 2016, p. 6. [Online]. Available: https://computingwithinlimits.org/2016/papers/a11-raghavan.pdf

[3] A. Borning, B. Friedman, and N. Logler, "The 'invisible' materiality of information technology," Communications of the ACM, vol. 63, no. 6, pp. 57– 64, 2020, doi: 10.1145/3360647.

[4] B. Sutherland, "Strategies for Degrowth Computing," presented at the LIMITS 2022 - Workshop on Computing Within Limits, Jun. 2022, p. 6. [Online]. Available: https://computingwithinlimits.org/2022/papers/limits22-final-Sutherland.pdf DOI:https://doi.org/10.1145/3401335.3401825

[5] R. R. Abbing, "'This is a solar-powered website, which means it sometimes goes offline': a design inquiry into degrowth and ICT," LIMITS Workshop on Computing within Limits, Jun. 2021, doi: 10.21428/bf6fb269.e78d19f6.

[6] T. Brain, A. Nathanson, and B. Piantella, "Solar Protocol: Exploring Energy-Centered Design," 2022, p. 7.

[7] BBC News, "Right to repair rules will extend lifespan of products, government says," BBC News, Jun. 30, 2021. Accessed: Mar. 31, 2023. [Online]. Available: https://www.bbc.com/news/business-57665593

[8] S. Ahmed, "Making Feminist Points," feministkilljoys, Sep. 11, 2013. https://web.archive.org/web/20230330150439/https://feministkilljoys.com/2013/09/11/making-feminist-points/ (accessed Mar. 27, 2023).

[9] F. Birol, "The Role of Critical Minerals in Clean Energy Transitions," International Energy Agency, Mar. 2022. p. 5 Accessed: Jan. 30, 2023. [Online]. Available: https://web.archive.org/web/20230321101012/https://iea.blob.core.windows.net/assets/ffd2a83b-8c30-4e9d-980a-52b6d9a86fdc/TheRoleofCriticalMineralsinCleanEnergyTransitions.pdf

[10] E. Manzini, "Design, Environment and Social Quality: From 'Existenzminimum' to 'Quality Maximum,'" Design Issues, vol. 10, no. 1, pp. 37–43, 1994, doi: 10.2307/1511653.

[11] E. Lemberger, "Elliott Lemberger - Senior Staff, Energy Experience Design - Tesla | LinkedIn," Apr. 01, 2023. https://www.linkedin.com/in/elliottlemberger (accessed Apr. 01, 2023).

[12] J. Law, Ed., A Sociology of Monsters: Essays on Power, Technology and Domination. London: Routledge, 1991. [Online]. Available: https://edisciplinas.usp.br/pluginfile.php/4621898/mod_resource/content/1/JAW%2C%20John.%20A%20Sociology%20of%20monsters.pdf

[13] F. Shunaman, "Can Dry Cells Be Recharged?," Popular Electronics, Jul. 1967.

[14] C. Carles and T. Ortiz, "Regen Box Project Funding Page," Ulule Crowdfunding Platform, Oct. 2016. https://web.archive.org/web/20210427093552/https://www.ulule.com/regenbox

[15] J. F. Randall, Designing indoor solar products: photovoltaic technologies for AES. Hoboken, N.J: J. Wiley & Sons, 2005.

[16] L. Hartmann, "MNT Reform: The Much More Personal Computer," Interface Critique Journal, vol. 1, no. Beyond UX, 2018, doi: https://doi.org/10.11588/ic.2018.1.44735.

[17] Fairphone, "Fairphone 4 - Sustainable. Long-lasting. Fair. | Fairphone," Mar. 25, 2023. https://web.archive.org/web/20230325234825/https://shop.fairphone.com/en/?ref=header (accessed Apr. 01, 2023).

[18] L. Suchman, "Anthropological Relocations and the Limits of Design," Annu. Rev. Anthropol., vol. 40, no. 1, pp. 1–18, Oct. 2011, doi: 10.1146/annurev.anthro.041608.105640.

[19] S. P. Shackleton and H. W. Purcell, "Relays in the Bell System," The Bell System Technical Journal, vol. 3, no. 1, pp. 1–42, Jan. 1924, doi: 10.1002/j.1538-7305.1924.tb01346.x.

[20] H. J. Becker and V. Ferry, "Low Voltage Electrolytic Capacitor," Jul. 23, 1957 [Online]. Available: https://patentimages.storage.googleapis.com/a2/f8/a9/b7d5c04a415c8b/US2800616.pdf

[21] L. Greenemeier, "The Dark Horse in the Race to Power Hybrid Cars," Scientific American, Aug. 28, 2007. Accessed: Apr. 01, 2023. [Online]. Available: https://web.archive.org/web/20230928213429/https://www.scientificamerican.com/article/the-dark-horse-in-race-to/

[22] Panasonic Corporation, "Electric Double Layer Capacitors | Panasonic Industrial [Devices," Dec. 03, 2021. Available: https://web.archive.org/web/20221203170154/https://na.industrial.panasonic.com/products/capacitors/electric-double-layer-capacitors-gold-capacitor (accessed Apr. 01, 2023).

[23] Eaton Electronics, "Technical Data 11043: HS/HSL Supercapacitors." Eaton Electronics, Nov. 2021. [Online]. Available: https://web.archive.org/web/20211103092444/https://www.eaton.com/content/d am/eaton/products/electronic-components/resources/data-sheet/eatonsupercapacitor-hybrid-cylindrical-cells-data-sheet.pdf